\documentclass[sigconf]{acmart}
\usepackage{multirow}
\usepackage{tcolorbox}
\usepackage{listings}
\usepackage{booktabs}
\usepackage{multirow}
\usepackage{graphicx}
\usepackage{makecell}
\newif\ifdraft
\drafttrue
\newcommand{\nb}[2]{
	{
		{\color{black}{
				\fbox{\bfseries\sffamily\scriptsize#1}
				{\sffamily$\triangleright~${\it\sffamily #2}$~\triangleleft$}
	}}}
}

\ifdraft
\newcommand\raula[1]{\nb{Raula}{\color{blue}#1}}
\newcommand\grex[1]{\nb{Gregorio}{\color{red}#1}}
\newcommand\chaiyong[1]{\nb{Chaiyong}{\color{magenta}#1}}
\newcommand\reviewer[1]{\nb{Reviewer}{\color{red}#1}}
\newcommand{\fixme}[1]{{\textcolor{red}{[FIXME] #1}}\xspace}

\else
\usepackage[disable]{todonotes}
\newcommand\raula[1]{}
\newcommand\grex[1]{}
\newcommand\chaiyong[1]{}
\newcommand\reviewer[1]{}
\newcommand{\fixme}[1]{}

\fi

\usepackage[inline]{enumitem}

\colorlet{punct}{red!60!black}
\definecolor{background}{HTML}{EEEEEE}
\definecolor{delim}{RGB}{20,105,176}
\colorlet{numb}{magenta!60!black}

\lstdefinelanguage{json}{
  basicstyle=\footnotesize\ttfamily,
  numbers=left,
  numberstyle=\scriptsize,
  stepnumber=1,
  numbersep=6pt,
  showstringspaces=false,
  breaklines=true,
  frame=lines,
  backgroundcolor=\color{background},
  literate=
   *{0}{{{\color{numb}0}}}{1}
   {1}{{{\color{numb}1}}}{1}
   {2}{{{\color{numb}2}}}{1}
   {3}{{{\color{numb}3}}}{1}
   {4}{{{\color{numb}4}}}{1}
   {5}{{{\color{numb}5}}}{1}
   {6}{{{\color{numb}6}}}{1}
   {7}{{{\color{numb}7}}}{1}
   {8}{{{\color{numb}8}}}{1}
   {9}{{{\color{numb}9}}}{1}
   {:}{{{\color{punct}{:}}}}{1}
   {,}{{{\color{punct}{,}}}}{1}
   {\{}{{{\color{delim}{\{}}}}{1}
   {\}}{{{\color{delim}{\}}}}}{1}
   {[}{{{\color{delim}{[}}}}{1}
   {]}{{{\color{delim}{]}}}}{1},
}

\lstdefinelanguage{bash}{
  basicstyle=\normalfont\ttfamily,
  numbers=left,
  numberstyle=\scriptsize,
  stepnumber=1,
  numbersep=8pt,
  showstringspaces=false,
  breaklines=true,
  frame=lines,
  backgroundcolor=\color{background}
}

\AtBeginDocument{%
 \providecommand\BibTeX{{%
  \normalfont B\kern-0.5em{\scshape i\kern-0.25em b}\kern-0.8em\TeX}}}

\setcopyright{acmcopyright}
\copyrightyear{2021}
\acmYear{2021}
\acmDOI{10.1145/1122445.1122456}

\acmConference[ICPC '22]{The 30th IEEE/ACM International Conference on Program Comprehension}{May 16--17, 2022}{Pittsburgh, USA}
\acmBooktitle{ICPC '22: The 30th IEEE/ACM International Conference on Program Comprehension, May 16--17, 2022, Pittsburgh, US}
\acmPrice{15.00}
\acmISBN{978-1-4503-XXXX-X/18/06}
\begin{document}

\title{\texttt{pycefr}: Python Competency Level through Code Analysis}
\author{Gregorio Robles}
\email{grex@gsyc.urjc.es}
\affiliation{%
 \institution{Universidad Rey Juan Carlos}
 \city{Madrid}
 \country{Spain}
}

\author{Raula Gaikovina Kula}
\email{raula-k@is.naist.jp}
\affiliation{%
 \institution{NAIST}
 \city{Nara}
 \country{Japan}
}

\author{Chaiyong Ragkhitwetsagul}
\email{chaiyong.rag@mahidol.edu}
\affiliation{%
 \institution{Faculty of ICT, Mahidol University}
 \city{Nakhon Pathom}
 \country{Thailand}
}

\author{Tattiya Sakulniwat}
\email{tattiya.sakul@gmail.com}
\affiliation{%
 \institution{NAIST}
 \city{Nara}
 \country{Japan}
}

\author{Kenichi Matsumoto}
\email{matumoto@is.naist.jp}
\affiliation{%
 \institution{NAIST}
 \city{Nara}
 \country{Japan}
}

\author{Jesus M. Gonzalez-Barahona}
\email{jgb@gsyc.urjc.es}
\affiliation{%
 \institution{Universidad Rey Juan Carlos}
 \city{Madrid}
 \country{Spain}
}

\begin{abstract}

Python is known to be a versatile language, well suited both for beginners and advanced users.
Some elements of the language are easier to understand than others: some are found in any kind of code, while some others are used only by experienced programmers. The use of these elements lead to different ways to code, depending on the experience with the language and the knowledge of its elements, the general programming competence and programming skills, etc.
In this paper, we present \texttt{pycefr}, a tool that detects the use of the different elements of the Python language, effectively measuring the level of Python proficiency required to comprehend and deal with a fragment of Python code.
Following the well-known Common European Framework of Reference for Languages (CEFR), widely used for natural languages, \texttt{pycefr} categorizes Python code in six levels, depending on the proficiency required to create and understand it.
We also discuss different use cases for \texttt{pycefr}: identifying code snippets that can be understood by developers with a certain proficiency, labeling code examples in online resources such as Stackoverflow and GitHub to suit them to a certain level of competency, helping in the onboarding process of new developers in Open Source Software projects, etc.
A video shows availability and usage of the tool: \url{https://tinyurl.com/ypdt3fwe}.
\end{abstract}



\maketitle
\renewcommand{\shortauthors}{Robles, Kula, Ragkhitwetsagul, et al.}

\section{Introduction}

Python is one of the most used languages nowadays, consistently ranked among the top programming languages during the past decade~\cite{IEEETopProg2021,TIOBETopProg2021}, with many top-tier IT companies using it. 
Python is known to be versatile, suitable for a wide range of people, from beginners who want to create small scripts to professional programmers.
It is also used by many developers not having a formal computer science education, as is for example the case of the BioPython collection~\cite{cock2009biopython}.
Much of science-related software is developed in Python, and much of it by non-IT professionals.

A live community, with its own \emph{culture}, has formed around Python.
An important part of this culture is about how to conceive solutions for certain problems, as is portrayed in the ``The Zen of Python''\footnote{https://www.python.org/dev/peps/pep-0020/}. One of its rules says:
{\em ``There should be one --and preferably only one-- obvious way to do it.''}
As a result, some solutions are considered more \emph{elegant} than others,
the most elegant being the most \emph{Pythonic}~\cite{alexandru2018usage}.

Our hypothesis, however, is that this is not completely true.
There are many different ways to program Python, and our experience is that the proficiency of the developer is key to understand certain ways (i.e., idioms) found in the Python language.
So, while we agree that creating more \emph{Pythonic} code should be the goal when learning Python,
less proficient developers could use other elements that do not require a deep knowledge of the language.


To validate our hypothesis, we have built \texttt{pycefr}, a tool that analyzes a given piece of code, reports the different Python elements found in it, and estimates the proficiency level required by a developer to understand it. Following the design of the Common European Framework of Reference for Languages (CEFR)~\cite{CEFR}, we characterize Python code in a scale of six proficiency levels. Mimicking the process of learning natural languages, we postulate that we start off with a few elements and easy structures, which with experience and learning we extend, acquiring more complex and advanced elements. 

We envision that our tool could be helpful in several scenarios: 
i) in Open Source Software (OSS) projects, to select proper reviewers and bug fixers, making onboarding of new developers~\cite{steinmacher2015systematic} easier;
ii) in educational environments, to set an upper bound of the programming skills required, in particular where programming is secondary (e.g., in computer networks class, where students have to create simple Python clients and servers, the important concepts are about networks and protocols, requiring just the minimum necessary Python skills);
iii) in social Q\&A sites, such as Stackoverflow, to understand the competency needed to comprehend a proposed answer
(although simply copying code from Stackoverflow might work, if it is not understood by the developer it may work in an unexpected way~\cite{Acar2016,Zhang2018,Ragkhitwetsagul2021}).

\section{A Framework for Language Proficiency}

The ``Common European Framework of Reference for Languages: Learning, Teaching, Assessment'', popularly known as CEFR in English~\cite{CEFR}, is a guide created by the Council of Europe to describe the performance of foreign natural language learners, mainly in Europe but it has also been increasingly adopted in other countries. Its main objective is to provide a method of learning, teaching and assessment that applies to all the natural languages of Europe. 
The CEFR intends as well to make it easier for educational institutions and employers to assess the language skills of applicants for admission to training or employment. 
CEFR defines the skills that citizens must have for each level, and offers thus a standardized and widely adopted way of expressing the proficiency in a language.

\begin{table}[tb]
\resizebox{.48\textwidth}{!}{
\begin{tabular}{cl}
\toprule
\textbf{Level group} & \textbf{Level} \\ 
\midrule
\multirow{2}{*}{\makecell{A\\Basic user}} & \makecell[l]{A1: Breakthrough or beginner} \\ \cmidrule{2-2} 
 & \makecell[l]{A2: Waystage or elementary} \\ 
 \midrule
\multirow{2}{*}{\makecell{B\\Independent user}} & \makecell[l]{B1: Threshold or intermediate} \\ \cmidrule{2-2} 
& \makecell[l]{B2: Vantage or upper intermediate}     \\ \midrule
\multirow{2}{*}{\makecell{C\\ Proficient user}} & \makecell[l]{C1: Effective operational proficiency or advanced} \\ \cmidrule{2-2} 
& \makecell[l]{C2: Mastery or proficiency} \\
\bottomrule
\end{tabular}
}
\caption{Reference levels in the Common European Framework of Reference for Languages (CEFR)}
\label{tab:cefr}
\vspace{-0.5cm}
\end{table}

CEFR has six reference levels (A1, A2, B1, B2, C1, C2) that are widely accepted as the European standard for classifying individual language skills, as can be seen from Table~\ref{tab:cefr}.
Levels in CEFR are incremental, in the sense that the abilities required for the B2 level implicitly require proficiency at lower levels (in this case, A1, A2 and B1)~\cite{north2007cefr}.
Several organizations have been created to serve as an umbrella for language schools and certification companies that claim to comply with the CEFR~\cite{martyniuk2011aligning}. For example, the European Association for Language Testing and Assessment (EALTA) is an initiative funded by the European Community that promotes the CEFR and best practices in the provision of professional language training~\cite{figueras2007cefr}. Initiatives at national level exist, too~\cite{deygers2018one,hancke2013exploring,kavakli2017applying}.

\section{Related Research \& Tools}

There are a few studies on \emph{Pythonicity} and its adoption in OSS projects.
Alexandru et al.~\cite{alexandru2018usage} study how the developers perceive the term \emph{Pythonic} and also propose a catalog of \emph{Pythonic} idioms and their usage in real-world projects. The study found that writing \emph{Pythonic} code can be an indicating factor of expertise of Python developers.
Sakulniwat et al.~\cite{sakulniwat2019visualizing} study the usage of the \texttt{with open} \emph{Pythonic} idiom over time in four Python projects. They found that the adoption of the \texttt{with open} increases over time in two projects, along with removals of its non-Pythonic counterpart.
Phan-udom et al.~\cite{phan2020teddy} propose an automated tool, called Teddy, that can recommend adoptions of \emph{Pythonic} idioms based on the code changes in GitHub pull requests of Python projects. 

Previous studies have proposed several approaches to measure developers' expertise or proficiency, for instance h-index for developers based on the data in OSS projects~\cite{capiluppi2012developing}, assessing candidates from their activities on social network~\cite{capiluppi2012assessing}, or the \texttt{Visual Resume} from past history and expertise from GitHub and Stack Overflow~\cite{sarma2016hiring}. 
It has been shown that developers' programming proficiency can affect their involvement in OSS communities.
When sharing knowledge in social Q\&A sites such as Stack Overflow, that has become widely used for OSS communities~\cite{vasilescu2014social}, asking a well-formulated question is beneficial for receiving useful answers~\cite{yang2014asking}. Moreover,
one of the challenges newcomers face when joining OSS projects is lack of technical expertise. Moreover, they also found difficulties in finding tasks that match with their programming proficiency~\cite{steinmacher2015systematic}.
Along the same line as our work, Vasilescu et al.~\cite{vasilescu2013babel} have applied the concept of linguistic diversity drawn from the field of social science to programming languages. They study the risks of programming languages to disappear or to become unmaintainable.


\begin{table}
	\centering
	\caption{Level Assigned to an excerpt of Python elements.}
	\begin{tabular}{clc}
		\toprule
		No & Python Element & Level \\
		\midrule
		1 & Print & A1 \\
		2 & If statement & A1 \\
		3 & List & A1 \\
		4 & Open function (files) & A2 \\
		5 & Nested list & A2 \\
		6 & List with a dictionary & B1 \\
		7 & Nested dictionary & B1 \\
		8 & with & B1 \\
		9 & List comprehension & B2 \\
		10 & \_\_dict\_\_ attribute & B2 \\
		11 & \_\_slots\_\_  & C1 \\
		12 & Generator function & C1 \\
		13 & Meta-class & C2 \\
		14 & Decorator class & C2 \\
		\bottomrule
	\end{tabular}
	\label{tab:python-elements}
\end{table}

\section{Building a Competency Framework}
\label{sub:building}

Our design assumes that, given a piece of code, the highest level of a Python element\footnote{Throughout this paper, we refer to a \emph{Python element} both as a Python element (e.g., a print statement, or as a construct (e.g., a list comprehension).} found will signal the proficiency that a Python developer has to have in order to master (i.e., understand, modify, fix, refactor) it.
In other words, given a code snippet where there are elements belonging to A1, A2 and B1, a developer should at least have a B1 level in order to completely understand and work with it. 

We have taken common Python elements and have assigned them to levels.
Table~\ref{tab:python-elements} shows an excerpt of the assignment table, where we show some elements for each level.
At this point, the complete list of identified elements is over 100 (which can be found in the replication package\footnote{https://doi.org/10.5281/zenodo.5577363}).
To create the list, we have studied several Python books~\cite{lutz2001programming,chun2001core,kuhlman2009python,pilgrim2009dive,summerfield2010programming,downey2012think}, have identified Python elements that are explained in them, and have written down the order in which they appear.
Two of the authors have taken this ordered list as input and have created a first version of the assignment matrix, for which given a Python element/construct returns its level.
This has been shared with the rest of the authors, and after discussion a certain agreement has been obtained.

We are still scientifically evaluating the assignments of Python constructs and elements to competency levels.
Further research is required for this.
We have noticed that the level might depend on the context, and even minor disagreements (i.e., small differences in the order and in the assignment, e.g., assigning B2 instead of B1) may exist.
In any case, it should be noticed that assignments can be easily changed in the configuration file of \texttt{pycefr}, without having to modify a line of source code of the tool.



\section{Detecting Python Proficiency Level}

Given a piece of code, \texttt{pycefr} will create the AST graph of the program, using Python3's \texttt{ast} standard library.
The program will then detect the individual Python elements according to the list (as explained in Section~\ref{sub:building}) and sum the number of times they appear.

 \begin{figure}[t]
   \centering
   \includegraphics[width=0.85\linewidth]{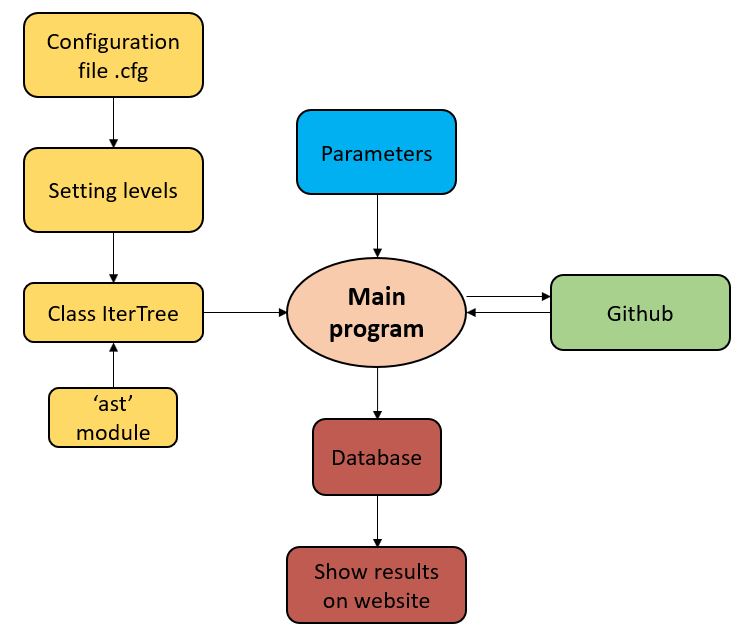}
   \caption{Schematic view of \texttt{pycefr}'s components and the flow of information.}
   \label{fig:schema}
 \end{figure}

Figure~\ref{fig:schema} offers a schematic view of how \texttt{pycefr} works.
As it can be seen, the main Python program receives the source (a GitHub username, a git repository, a folder or a file) as a parameter.
In the case of getting a GitHub username, the tool uses the GitHub API to retrieve the list of repositories of that user where Python is the main language and stores their git URL in a list.
Then it traverses this list, cloning those git repositories.
For each repository, its filetree is iterated, and all files with a \texttt{.py} extension are analyzed, which means that its AST tree is generated.
The Python elements found are then stored together with the information of where they have been found (file and line) in the database.

\begin{figure}
    \centering
    \includegraphics[width=0.7\linewidth]{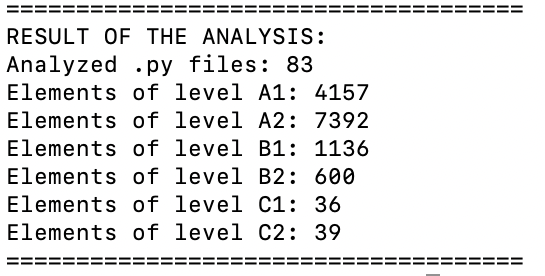}
    \caption{Output of the analysis of the Requests GitHub repository as provided by \texttt{pycefr}.}
    \label{fig:result_repourl}
\end{figure}



Figure~\ref{fig:result_repourl} shows a screenshot example of the user interface of \texttt{pycefr}.
In addition to the command-line output, \texttt{pycefr} will return the results in CSV and JSON format.
The output can be in either CSV or JSON file that contains following information: (i) Python element, line of code of the occurrence, and (ii) the level of proficiency.

\begin{figure}
  \centering
\begin{lstlisting}[language=json,firstnumber=1]
{
 "tools": {
  "clang_format_utils.py": [
   {
    "Class": "Simple Dictionary",
    "Start Line": "16",
    "End Line": "19",
    "Displacement": "21",
    "Level": "A2"
   },
   {
    "Class": "Print",
    "Start Line": "70",
    "End Line": "70",
    "Displacement": "4",
    "Level": "A1"
   },
   ...
}
\end{lstlisting}
 \caption{JSON output of a \texttt{pycefr} analysis (excerpt).} 
\label{fig:json}
\end{figure}

Figure~\ref{fig:json} shows the output of \texttt{pycefr} on the \texttt{clang\_format\_utils.py} file, taken from the PyTorch project. 
Furthermore, \texttt{pycefr} can summarize the analysis as an aggregation of all proficiency levels for that given code.
The JSON output is used as well as a simple database for creating of a set of web pages using HTML, JavaScript and CSS to offer results in a more visual manner to the user.




\section{Application}
\label{sec:application}

We have tested \texttt{pycefr} in a computer networks subject that two of the authors of the paper teach at university level (with 33 students in the 2019/2020 academic year).
Students who enroll in this subject have previous knowledge of programming (in Pascal, Ada and some of them in Java), but no Python knowledge.
That is why a crash course on Python is given for 8 hours.
We decided to teach Python up to level B1 (included), and removed Python elements that had been taught in previous years such as list comprehensions, which we have considered as B2.

 \begin{figure}
   \centering
   \includegraphics[width=\linewidth]{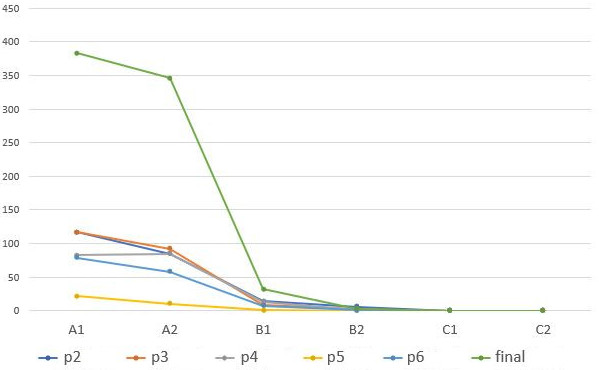}
   \caption{Number of elements per level found in student assignments.}
   \label{fig:comparison}
 \end{figure}

During the course, students have to submit 5 small assignments (from p2 to p6) and a large final assignment. 
All the assignments are submitted to a git repository.
Figure~\ref{fig:comparison} is the result on analyzing the repositories of the assignments by a student, which is representative of what we have found when analyzing them with \texttt{pycefr}.
As it can be seen, most of the Python elements used by the student are in the A levels. 
There are as well some elements with B1 level and even some of them B2 (14 in total).
No elements of the C levels were found. 
Inspecting which B2 elements showed up, we found two of them: a) the \emph{if \_\_name\_\_ == '\_\_main\_\_':} is used to execute some code only if the file was run directly, and not imported, i.e., when a Python program runs as a main program, and b) dictionaries of lists of dictionaries, which is a complex data element.
In the first case, we noted that we had not presented it in our Python crash course, but that we introduced it later in our lessons.
This points out to an incorrect level assignment, as probably it should belong to Levels A2 or B1, but we had put it as B2. The reason is because in the text books this element, even if used before, is really explained much later.
In the second case, we have shown dictionaries and lists, and their combination in our crash course, but not such a complex element.
However, the student herself realized that this is possible and introduced it in her code.

When analyzing other student's projects, we found other B2 (list comprehensions and lambda functions) and even C1 (decorators) elements in their code. 
We asked the authoring students about them and they told us that they had copied it after doing an Internet search (most of them from Stack Overflow although others did not remember).
All in all, our preliminary experience shows that we achieved to design assignments that students could do with the Python level they had, which was our goal.
Since the course is about computer networks, our objective was not about teaching students to learn more Python during the course but to ascertain that the Python of the crash course is enough for the whole course.

\section{Other Potential Applications}
\label{sec:other}

\textbf{Choosing Example Code Snippets}.
In books, tutorials and lessons, code snippets are often shown. These example code snippets may also be used as a starting point to extend the implementation. To make sure developers can fully understand the examples being shown, appropriate code snippets based on the reported level of proficiency can be chosen with the \texttt{pycefr} tool. For example, in a fundamentals of programming class using Python, educators can limit the level of code examples to only A2 or B1. On the other hand, for a more advanced course like data science or machine learning, code examples could include functional programming fragments (mostly labeled as B2).

\textbf{Increase the Understanding of Stack Overflow Code Snippets.}
The \texttt{pycefr} could be used to evaluate the level of code snippets in a Stack Overflow post including the question and the answers and add a label of Python proficiency level of the answers. This could enable developers with limited knowledge in Python code, such as the ones with B1 or below, to browse through the answers and focus on the ones that match best with their proficiency level. This may help developers to not only copying and using the code as-is, but also understanding the code that they reuse. In addition, it has been found that a number of Stack Overflow code snippets are problematic~~\cite{Acar2016,Zhang2018,Ragkhitwetsagul2021}. This may also help developers to spot issues in such code since they can fully understand them and make informed decisions whether to reuse it or not. 


\textbf{Onboarding OSS Projects.}
The \texttt{pycefr} tool can be used to encourage contributions to OSS projects. For example, newcomers can work on the components that have an average proficiency level lower than or equal to their proficiency level. Moreover, if the proficiency level is applied to the code review process in OSS projects, newcomers can choose which reviews they are capable of participating in. By knowing the level of Python proficiency that is required to fix a bug, software projects could attract newcomers more easily, as bug fixes that correspond to lower levels of Python proficiency could be taken by them. 


\textbf{Application to other programming languages.}
Beyond the impact on Python, we think that if our tool shows to have promising uses, these ideas could be transferred to other programming languages and platforms.
So, we argue that assigning levels for programming elements can be done in many (if not all) other languages, and that similar strategies as the ones presented above could be used in projects implemented in those languages, too.




\section{Roadmap and Future Outlook}
\label{sec:conclusion}
We now present limitations and the future plans for improvement.

\textbf{Limitations.}
At current writing, \texttt{pycefr} currently supports only Python 3. 
Although a large number of Python 2 programs are available~\cite{devto2020versionshare}, we focus on Python 3, which is fully-supported version and future-proof and aruge that Python 2 has a decreased adoption~\cite{jetbrain2020devsurvey}.
\texttt{pycefr} is mainly suited for software comprehension purposes, e.g., to identify the level of competency that a developer requires to perform a task, and is not intended, as an assessment tool because it can be easily circumvented. 
A simple example would be creating a function that is never called and that contains C1 and C2 elements.
CEFR has been created to evaluate the competence with a foreign natural language. 
Although it has been shown that programming languages share similar characteristics with natural languages~\cite{Hindle2012}, they may not be always equivalent.

\textbf{Future Work.}
Since competency is a sliding scale, we envision incremental re-calibration for improve \texttt{pycefr} proficiency accuracy. 
From the software comprehension perspective, \texttt{pycefr} offers the possibility to ensure that the code shown or given to developers is within their level of Python competency.
This should also open up to new methodologies on how developer comprehend of the skill of programming, and how it can be applied in practical scenarios as mentioned in the paper.

\section*{Acknowledgements}

GR acknowledges the support of the Madrid Regional Government through the e-Madrid-CM Project under Grant S2018/TCS-4307, co-funded by the European Structural Funds (FSE and FEDER).
RK, TS, and KM are supported by the Japanese Society for the Promotion of Science (JSPS) KAKENHI Grant Numbers JP20K19774 and JP20H05706.

\bibliographystyle{ACM-Reference-Format}
\bibliography{references}

\end{document}